\documentclass[12pt, prd, reprint, amsmath, amssymb, aps, superscriptaddress]{revtex4-1}

\usepackage{appendix}
\usepackage{color}
\usepackage[dvipsnames]{xcolor}
\usepackage{graphicx}
\usepackage[citecolor=blue, linkcolor=blue, urlcolor=blue, colorlinks=true]{hyperref}

\def\beq{\begin{eqnarray}}
\def\eeq{\end{eqnarray}}
\def\beqa{\begin{eqnarray}}
\def\eeqa{\end{eqnarray}}

\definecolor{darkred}{rgb}{.743,0,0}

\begin{document}
\title{Ph.D. in physics as a hurdle race, and the ``glass hurdles" for women}

\author{Meytal Eran-Jona$^{1a}$, and Yosef Nir$^{2b}$}
\affiliation{$^1$Feinberg Graduate School, Weizmann Institute of Science, Rehovot, Israel 7610001\\
$^2$Department of Particle Physics and Astrophysics, Weizmann Institute of Science, Rehovot, Israel 7610001}
\email{$^a$meytal.jona@weizmann.ac.il; $^b$yosef.nir@weizmann.ac.il}

\begin{abstract}
On their way to an academic career in physics, Ph.D. students have to overcome difficulties at many levels. Beyond the intellectual challenge, there are also psychological, social and economic barriers. We studied the difficulties experienced by physics Ph.D. students in the Israeli universities, with special attention to gender-related issues. Among the hurdles that are much more significant for women than for men -- that we call ``the glass hurdles" -- we find gender-related discrimination, sexual harassment, physiological and psychological health issues, and challenges related to pregnancy and parenthood. We make recommendations for ways to confront and remove these barriers in order to provide female physicists with an equal opportunity to succeed.
\end{abstract}

\maketitle

\section{Introduction}
The proportion of women who acquire higher education has been increasing steadily from the 1950's. Nowadays, they constitute a majority among undergraduate and graduate students in many disciplines. In light of the above, the lack of females among students and academic staff in Science, Technology, Engineering and Mathematics (STEM fields), has been the focal point of research and action in Western democracies for the last two decades.

In this study, we focus on physics. Physics as a knowledge field is characterized by consistent gender imbalance in most countries. For example, in the US, of all sciences, physics has the lowest female participation rates: 21\% of undergraduate students, 20\% of Ph.D. students and 16\% of faculty members in physics departments \cite{porter2019}. These percentages, which have not changed significantly for over a decade, are similar in the UK, with some European countries having even poorer ratios \cite{cochran2019,blue2018}.

The contemporary picture in Israel is even worse. Women constitute 16\% of all Physics B.Sc., M.Sc. and Ph.D. students, and only 6\% of the academic staff within Israeli universities \cite{jona2019}. In life sciences, the picture is different. For instance, in medicine, women constitute 69\% of all Ph.D. graduates and 35\% of the academic staff, and in biology, women constitute 58\% of all Ph.D. graduates and 30\% of the academic staff.

Doctoral students are an understudied group that is of particular interest in the context of investigating the gender gap in STEM fields. Previous research reveals a high attrition rate for women before and during the postdoctoral studies, a key period towards academic careers, where the numbers of women decrease dramatically \cite{bostwick2018,goulden2009,gofen2011,carmi2011}. Therefore, it is interesting to examine how women experience doctoral studies: Do they face different difficulties than men? Does this experience explain, even if partially, the lower percentage of women who choose to pursue an academic career in physics?
We used mixed methods methodologies to answer these questions, combining nationwide representative survey of all physics Ph.D. students in Israel with in-depth interviews with Ph.D. female students.

\section{Literature review}
In the following we review the literature regarding obstacles and difficulties that graduate students in physics face along their studies, with some references to other STEM fields and to research on academic staff. In particular, we focus on gender differences and on the challenges for female physicists within academy.

\subsection{The culture of physics as a male dominated field}
Traweek was one of the first social scientists that focused her research on the physics community \cite{traweek1988}. She studied the culture of physics in high energy laboratories, and found a community that is based on intense competition. The available resources are, however, limited and are often distributed on the basis of biased decisions and social connections. She found people who see physics as the pinnacle of rationality, empty of emotion, and void of human influence.

The culture of physics as a competitive and male dominated field that Traweek observed may imperil women's participation in the field. If women are seen as contrary to science - particularly the "fundamental" and "objective" science of physics - then they may be immediately seen by the gatekeepers of science and community members as being unfit. Physics is associated with rationality, objectivity, and logic, features that have been historically associated with masculinity \cite{harding1991}.

Three decades after Traweek's work, research into diverse physics sub-fields in various countries reveals that the culture of physics as a male dominated field persists. The masculine working-culture, alongside growing demands for competitiveness and dedication to work, pose an obstacle for women. Moreover, reconciling work and private life becomes more difficult in a more precarious model of career demanding mobility, and brings new challenges for partners in dual career couples \cite{sekula2018}. Women in the field of physics have not only to navigate the masculine norms of the discipline, but also to negotiate the limited possible identities for female in physics \cite{gonsalves2016}.

Discussing the power relation within academy as a whole, Bagihole and Goode \cite{bagilhole2001} claim that there is a fixed, standard academic career model that is not gender-neutral, but is based on a masculine model anchored in hegemonic masculine culture and a patriarchal support system. In a study about MIT senior female faculty, it was found that the ``ideal" perfect academic is one who gives total priority to work and has no outside interests and responsibilities \cite{bailyn2003}. What we see in physics is a vicious circle, whereas the absence of women-physicists generates reluctance among women to make the effort to fit into a field where they will be a marginal minority. Physics as a masculine field is one of the persisting ``castles" of gender imbalance.

The perception of a profession as male or female is also influenced by the extent to which an occupation allows or does not allow to combine family life with a career. In a study of women who completed their Ph.D. in STEM fields with excellent grades, this component was found to play a significant role in the decision-making of  whether to pursue an academic career in science \cite{gofen2011}.

Lamont and Molnar \cite{lamont2002} explain the preservation of segregation between women and men via the term ``boundaries". Boundaries are complex structures - physical, social, and psychological - that produce commonalities and differences between women and men, and in turn shape and structure the behavior and attitudes of each gender \cite{gerson1985}. Social boundaries are used to distinguish between women and men in the workplace. Thus, male employees are perceived as more competent than female employees. Those who violate gender boundaries and accepted norms, such as the accepted norm of dedicating yourself and all your time to work, may suffer from stigma and punishment in the workplace \cite{epstein2004}. Looking at physics departments worldwide, it seems that there are clear boundaries that prevent many talented women from choosing a career in physics, and that those boundaries are closely related to the culture of physics as a masculine arena.

\subsection{Mental health within faculty and Ph.D. students in physics}
Research on undergraduate and graduate students shows higher rates of mental health issues among students compared to the overall population. Literature review of mental health in research environment, mainly in the UK \cite{guthrie2017}, indicates that Ph.D. students face mental health issues. The proportions of both university staff and postgraduate students with a risk of having or developing a mental health problem, based on self-reported evidence, were generally higher than for other working populations. Moreover, large proportions ($>40\%$) of postgraduate students in the UK report symptoms of depression, emotion or stress-related problems, or high levels of stress. The main factors associated with the development of depression and other common mental health problems in Ph.D. students are high levels of work demands; the pressure to publish and win grants in highly competitive environments; job insecurity;  work-life conflict; low job control; poor support from the supervisor and exclusion from decision making. Believing that Ph.D. work is valuable for one's future career helps reduce stress, as does confidence in one's own research abilities.

Furthermore, gender was found as the key personal factor that contributes to mental health outcomes in the research workplace. Women report more exposure to stress than men. Moreover, they also report greater challenges around work-life balance. The results on whether age affects mental health were inconclusive, partly because age is often difficult to disentangle from the discussions about rank and seniority. Other factors such as disability, sexuality and minority status were mentioned in a small number of articles, which indicate that these personal factors generally increase stress \cite{guthrie2017}.

\subsection{Work-life conflict in academy}
Work-life conflict is a source of stress related to workload. A survey conducted among all active members of University and College Union (UCU) \cite{footnote:UCU} and reported in Ref.~\cite{kinman2013} shows that work demands are the strongest predictor of work-life conflict. In that survey, the majority of respondents reported that their ideal level of work-life separation would be greater than what they experienced at the time of reporting \cite{kinman2013}. Tytherleigh {\it et al.} \cite{tytherleigh2005} also found that work-life conflict and work overload were sources of stress for higher education staff, but that the stress levels associated with these stressors were lower than for individuals working in other areas \cite{guthrie2017}.

In a Belgian study, work-life conflict was identified Among Ph.D. students as the most important predictor of mental health problems, followed by work demands \cite{levecque2017}. This factor was also identified as important in a UK study of Ph.D. students, which found that ``having a high workload that impacts on your private life" was a bothersome issue for respondents \cite{hargreaves2014}.

Research focused on a worldwide sample of physicists indicates that by an almost two-to-one margin, women are more likely than men to say that becoming a parent significantly affected their work in various ways. Women were most likely to report changing their schedules, spending less time at work, and becoming more efficient (when having children). Those findings echo results from the first two IUPAP surveys, in which women physicists reported that having children forced them to become more efficient because they had to leave their laboratory or office in time to pick up young children from childcare. The survey also asked respondents whether their employers had assigned less challenging work to them when they became parents. The majority of physicists did not report a change. Still, women were more likely than men to report being given less challenging work, and the difference was statistically significant \cite{ivie2012}.

\subsection{Gender based discrimination in physics}
Sexism occurs when men are believed to be superior to women. It is thought to be one of the reasons for women's under-representation in physics. The issue of sexism in physics and astronomy has not been thoroughly explored in the literature and there is currently neither much evidence for it, nor even clear language to discuss it. Ref.~\cite{barthelemy2016} is one of the few relevant research projects. It is based on interviews with women in graduate physics and astronomy programs about their individual experiences of sexism. Although a minority of the women interviewed did not report experiencing sexual discrimination, the majority experienced subtle insults and microaggressions. Other participants also experienced more traditional hostile sexism in the form of sexual harassment, gender role stereotypes, and overt discouragement.

Microaggression is a term describing a subtle form of gender bias. Among the dominating themes or forms, one finds sexual objectification, second-class citizenship, use of sexist language, assumption of inferiority, restrictive gender roles, invisibility and sexist jokes, as well as denial of the reality of sexism \cite{sue2010,barthelemy2016}. It is argued that microaggressions ``act upon women in several ways, by reiterating the social view that men are more valued than women, by reinforcing traditional stereotypes about proper gender roles, and by contributing to violence toward women by objectifying and sexualizing them" \cite{barthelemy2016}. Therefore, the consequences of microaggressions may be as severe as those of overt sexism.

Research has found that female physicists, including graduate students and faculty, frequently encounter microaggressions. Interviews with physicists (44 female and 22 male) from twelve  research institutions in eight European countries, indicate that women in physics face various forms of microaggressions, including assumption of inferiority, restrictive gender roles, sexist jokes, invisibility and sexual objectification. Furthermore, female physicists more often declare being unequally treated in their workplace than their male counterparts do. The significance of microaggressions is that it signals deprecation of professional position of female physicists, evoke negative emotions in women and their accumulation may contribute to women leaving science \cite{sekula2018}.

\subsection{Sexual harassment in physics}
Sexual harassment is a form of gender discrimination. Broadly defined, sexual harassment is unwelcome or inappropriate behavior of a sexual nature that creates an uncomfortable or hostile environment. It comes in various forms, both subtle and overt. In the  study on sexual harassment in physics, Ref.~\cite{aycock2019} considers three specific types. First, ``sexist gender harassment" describes hostile or insulting remarks and actions based on one's gender, such as saying that women cannot do physics. Second, ``sexual gender harassment" refers to sexual remarks or conduct, like commenting on the shape of a woman's body. Third, ``unwanted sexual attention", such as requests for sexual favors or unwanted touching.

The data regarding the extent of sexual harassment in women vary. Some studies indicate that sexual harassment affects the majority of women in science, technology, engineering, mathematics, and medicine (STEMM) \cite{clancy2014}.

Research has shown that women in male dominated occupations are at greater risk of being sexually harassed, and that these experiences increase job turnover intentions and withdrawal from work. There are few indications for sexual harassment in physics specifically. A survey of undergraduate women physics students has shown that approximately three quarters (74\%) of survey respondents experienced at least one type of sexual harassment \cite{aycock2019}. It was also found that certain types of sexual harassment in physics predict a negative sense of belonging and exacerbate the imposter phenomenon \cite{aycock2019}.

These finding are not surprising, given previous studies showing that experiencing sexual harassment (in general) increases a woman's likelihood of leaving a STEM career \cite{johnson2018}. For those women who do stick with their field, harassment hurts their career, economic standing, and well-being \cite{howe2016}. In short, unchecked harassment creates a drain on talent through lost work, lost ideas, and lost people \cite{libarkin2019}.

\section{Data collection and analysis}
The research is based on mixed research methods, combining a nationwide survey of physics Ph.D. students and interviews. The research methods are  embedded within feminist research approaches and theories that provide framework and tools for looking into women's lives \cite{reinharz1992,devault1999,krumer2014}.

Research tools: The research team has compiled an online self-administrated questionnaire that was sent directly via e-mail to university physics Ph.D. students. The questionnaire included 107 questions, of which 6 were open ended questions. The questionnaire contains questions regarding the following topics: students' socio-demographic background, academic study track, attitudes regarding the academic environment, success indicators, combining family and studies, future employment expectations and intentions, desire to have an academic career, considerations in favor of and against postdoctoral studies, and aspects of discrimination and sexual harassment during academic studies.

Sample and Sampling: The deans of physics at the six research universities \cite{footnote:univIL} contacted all students ($N=404$) at the institution to answer the questionnaire. The research team made a special effort to encourage all students to respond, and a few reminders were sent. Respondents ($n=267$) accounted for 66\% of the population, of which 60 were women and 207 men. The research team made a special effort to raise women's response rates because of the small number of the population and the researchers' interest in this group. These efforts reached 94\% ($N = 64$, $n = 60$).

Maximum margin of errors: the maximum margin of error for the entire population is 3.6\%, for women 3.2\%, and for men 4.3\%. Due to the over-representation of women in the sample, the total number of students was weighted by gender, the data for the entire sample included in the paper are weighted.

The survey findings and its interpretation are supported by findings from a qualitative research that we conducted concurrently among physics students in Israel. The research team conducted 25 in-depth, face-to-face interviews with young women, all doctoral students in physics studying in Israeli universities. The interviews were conducted during 2017-2018. Most of the interviewees were in their late twenties or early thirties (age range was 26-36). The interviews were recorded and transcribed. Data analysis was done using the conventional methods of discourse analysis \cite{denzin1998} using the ATLAS.ti software. The main findings of the interviews concern the considerations taken by female Ph.D. students when deciding whether or not to continue for a post-doctoral positions, and are beyond the scope of this paper \cite{nir2020}.

\section{Findings}
One of the main goals of the research was to find out what are the main difficulties with which the students are struggling during their studies. To achieve this goal, we asked an open broad question: ``{\it If a close friend were to consult with you about Ph.D. studies in physics, which difficulties would you present to them?}" The indirect formulation of this question was aimed to examine what, in the students' view, are the main difficulties, independent of whether they experienced them in person. Based on qualitative analysis, we find three main areas of difficulties:
\begin{itemize}
\item Professional: Physics studies are difficult, intensive and frustrating. They involve a competitive environment. Success strongly depends on the Ph.D. advisor.
\item Economic: Salary (in fact, fellowship, which does not include social benefits) is low and unsatisfactory. Consequently it is difficult to provide for a family. Furthermore, there is a significant job insecurity.
\item Personal: The years of studies are often characterized by loneliness, uncertainties, and emotional difficulties.
\end{itemize}

We further asked about the personal experiences. In fact, the difficulties described above were personally experienced by many of the respondents, women and men alike. Yet, on some issues, we found significant differences between women and men. We now focus on these gender-related differences.

\begin{figure}[h]
 \begin{center}
  \includegraphics[width=0.99\linewidth]{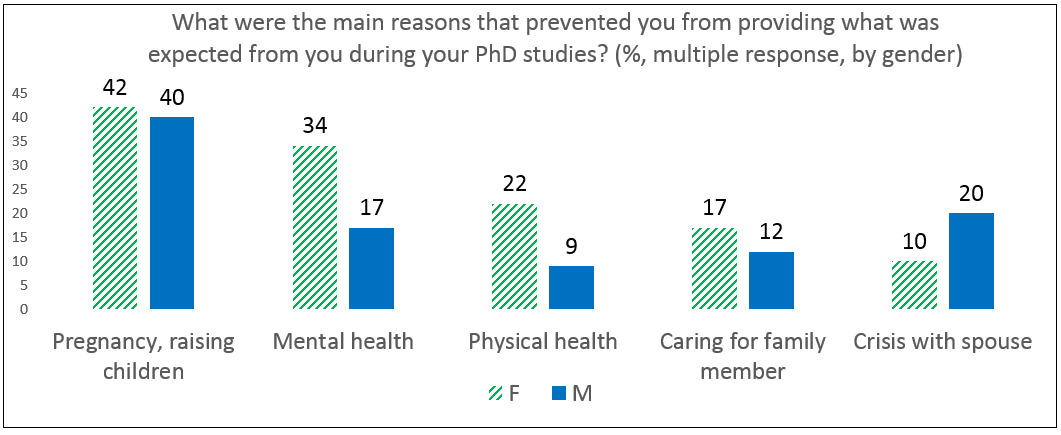}
  \caption{What were the main reasons that prevented you from providing what was expected from you during your Ph.D. studies? Data are presented by percentage, with patterned-green-left (solid-blue-right) column corresponding to female (male) Ph.D. students.
  }
  \label{fig:diff}
 \end{center}
\end{figure}

We asked: ``{\it During your Ph.D. studies, did you experience a period when it was difficult for you to provide what was professionally expected of you?}" Of the women, 71\% answered positively, compared to 63\% of men. When asked a closed question about the reasons for this difficulty (see Fig. \ref{fig:diff} for details), a much larger percentage of women (39\%) compared to men (23\%) mentioned health-related problems, both physiological and psychological.

In order to understand the significance of these data, we would like to refer to the uniqueness of the Israeli context, which has certain special characteristics. The Israeli society is very familial. Israelis marry on average at a relatively young age, and have more children than in other Western societies. Furthermore, because of the compulsory military service (2-3 years), Israeli students are older on average than their peers. Thus, in Israel, most of the Ph.D. students have already a spouse. Among the survey respondents, 70\% are married or in relationship, and 40\% are already parents. 95\% of mothers and 86\% of fathers stated that becoming parents affected their way of studying. A large majority of them (73\%) reduced the time spent on studies and research, and for a large fraction (34\%) this led to a reduced rate of progress in their research. Women mentioned more frequently than men that they learned to make their schedule more flexible (60\% {\it vs.} 48\%) and to be more efficient and productive (40\% {\it vs.} 27\%).

Because women are the ones to give birth, to breastfeed and to take care of the newborn during parental leave (the Hebrew term translates, somewhat ironically, to ``birth vacation"), we assumed that combining pregnancy and parenthood with studies should be much more challenging to them. Indeed, when we asked about the parental leave, 69\% of the mother-students took a four-month leave (which is the standard by law). In contrast, 58\% of father-students took no leave, and only 16\% of men took a leave longer than a month. We conclude that giving birth translates into a substantial delay in the Ph.D. progress for mothers, creating a significant gap compared to their male colleagues.

\begin{figure}
\centering
\includegraphics[width=0.96\linewidth]{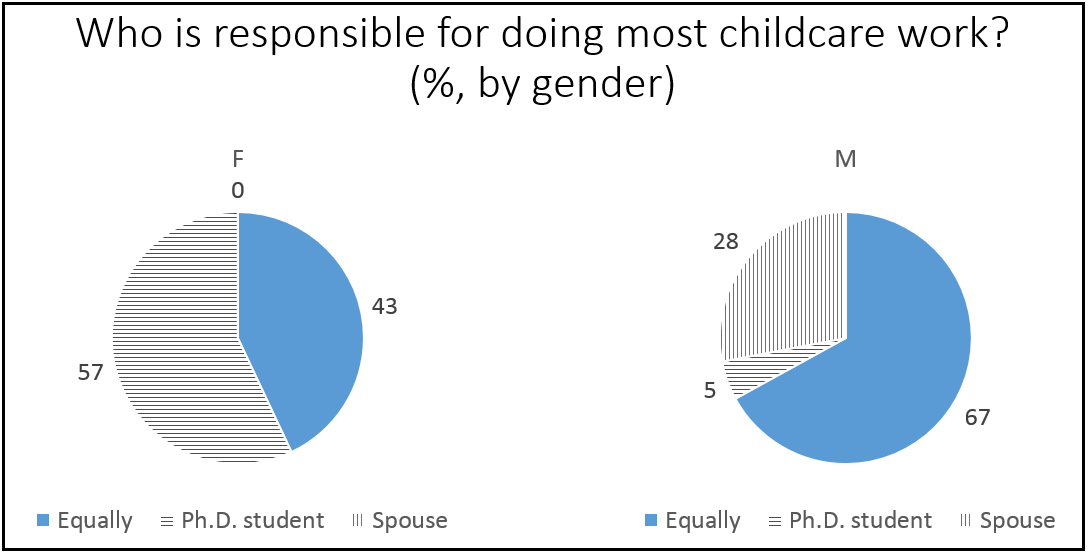} 
\caption{Who is responsible for doing most childcare work?
Data are presented by percentage, with left (right) pie corresponding to female (male) Ph.D. students.}
\label{fig:child}
\end{figure}

Another aspect of significant gender difference, familiar from studies around the world, arises from looking into the private sphere of the families of the Ph.D. students. We asked: ``{\it Who is responsible for most childcare work?}" Of the male students, 67\% responded that they and their spouses carry the load equally, and only 5\% responded that the responsibility lies mainly on themselves. In contrast, of the female students, 43\% responded that they and their spouses carry the load equally, and the other 57\% responded that the responsibility lies mainly on themselves. {\it Not even one of the female students said that her spouse is the main caregiver for the children.} See Fig. \ref{fig:child} for details.

\begin{figure}
\centering
\includegraphics[width=0.96\linewidth]{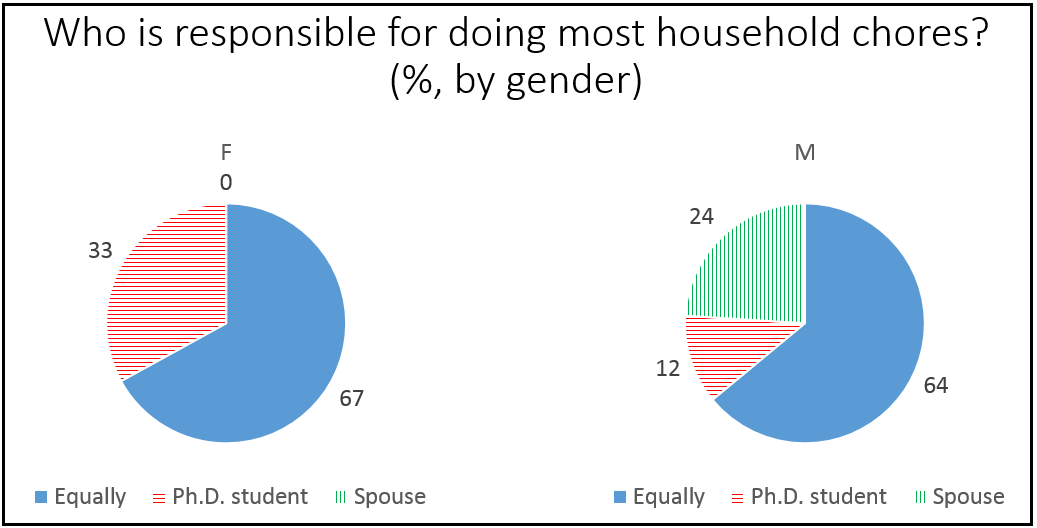}
\caption{Who is responsible for doing most
household chores?
Data are presented by percentage, with left (right) pie corresponding to female (male) Ph.D. students.}
\label{fig:house}
\end{figure}

A complimentary aspect is that of household chores. Again, we asked: ``{\it Who is responsible for doing most household chores?}" The emerging picture is quite similar to that of childcare. We observe again an ideology (and probably practice) of equality between the couple, with 67\% of female students and 64\% of male students stating an equal sharing of the household responsibilities. Yet, again, {\it not even one of the female students said that her spouse is responsible for doing most chores.} See Fig. \ref{fig:house} for details.


\begin{figure}
\centering
\includegraphics[width=0.96\linewidth]{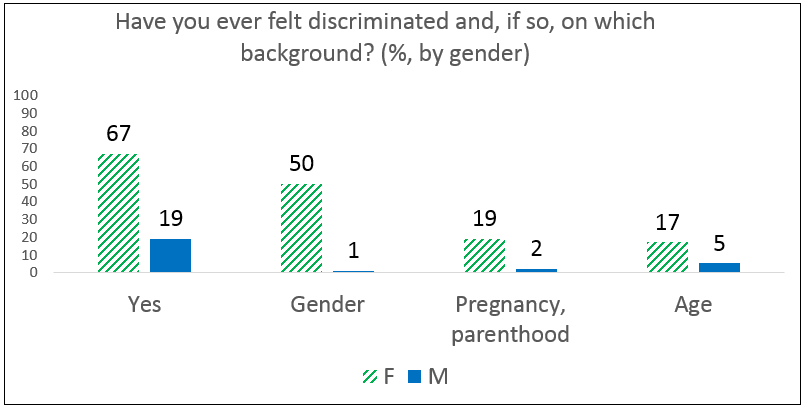} 
\caption{During your academic career, have you felt discriminated and, if so, on which background?
Data are presented by percentage, with patterned-green-left (solid-blue-right) column corresponding to female (male) Ph.D. students.}
\label{fig:discrimination}
\end{figure}

Two final issues where we identify gender-related differences are those of discrimination and of sexual harassment. We aimed to examine the issue of discrimination on a broad variety of possible backgrounds. We asked: ``{\it Have you ever felt discriminated against during your studies?}" The gender difference here is very significant, as 67\% of women {\it vs.} only 19\% of men have experienced discrimination. When asked on the grounds for the discrimination, 50\% of women reported ``gender" and 19\% reported ``pregnancy" or ``parenthood", while for men these issues were rarely mentioned. Another aspect of discrimination that was reported is age (17\% of women, 5\% of men). Other aspects - ethnic origin, religion and social status - were mentioned by only very few respondents. See Fig. \ref{fig:discrimination} for details.

\begin{figure}
\centering
\includegraphics[width=0.96\linewidth]{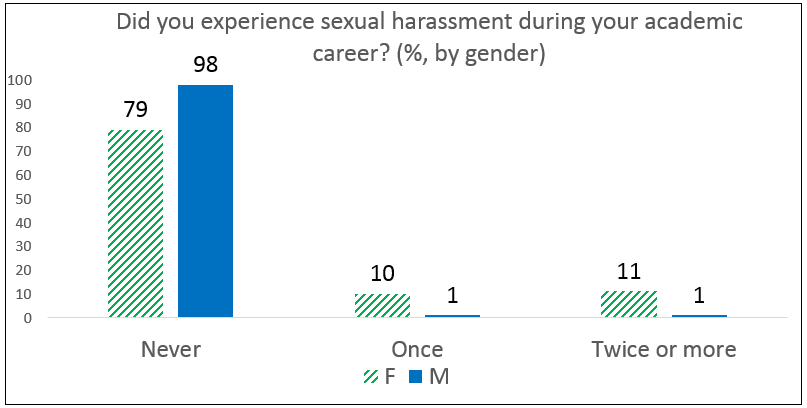}
\caption{During your academic career,
did you experience sexual harassment?
Data are presented by percentage, with patterned-green-left (solid-blue-right) column corresponding to female (male) Ph.D. students.}
\label{fig:harassment}
\end{figure}

On the issue of sexual harassment, we asked: ``{\it Did you experience sexual harassment during your academic studies?}" To avoid any ambiguity, we provided the Israeli legal definition of sexual harassment\cite{footnote:shIL}. One of every five women (21\%), but only 2\% of men, reported that they experienced sexual harassment during their studies. (See Fig. \ref{fig:harassment} for details.) Among these, half of the women were harassed twice or more. Only a minority of the women answered a question: ``{\it By whom were you harassed?}" The answers included student-colleagues, technicians and lecturers, but none pointed out the Ph.D. advisor.


\section{Discussion}
This research provides a broad, quantitative and representative examination of the challenges facing physics Ph.D. students. Our findings confirm in part the findings of previous studies. Its uniqueness is in referring to a large variety of difficulties, starting with academic requirements, via difficulties related to family and parenthood, and up to discrimination and harassment. Furthermore, in all of these issues we examine gender-related differences, and thus we learn about the unique experience of women physicists as a minority in a male-dominated field.

The Ph.D. students view the physics study course as one that is demanding, intensive and difficult, and which requires overcoming professional, economic and emotional hurdles. In the academic aspect, there are many challenges that all students face, including the need to invest a great deal of time in their research, workload, frustration and competitiveness. Because the curriculum demands do not allow for paid work, the students depend on a modest scholarship, which is significantly lower than their earning capacity in the labor market. The economic difficulty is exacerbated and stands out as a significant hurdle for married students. Moreover, coping with all of these also produces mental difficulties that manifest themselves in stress, anxiety, a continuing sense of uncertainty and mental distress.

In addition to difficulties experienced by all students, women have to face additional hurdles, that we call ``glass hurdles", of several types. First, women suffer more from problems in psychological and physiological health. Second, women face challenges related to pregnancy, giving birth and motherhood. Although the transition to parenting poses a significant challenge for both women and men, pregnancy and childbirth halt women course of study much more. In addition, after giving birth, they carry a heavier burden of childcare and the household chores compared to their spouses. Third, the findings show that while most women experienced discrimination during their studies, mainly on the grounds of gender, parenting and family, most men did not experience discrimination at all. Furthermore, the findings indicate an alarming picture that one in five women experienced sexual harassment during their studies (compared to a marginal rate among men). In this context, it is important to note that the literature points to serious and long-term effects of sexual harassment on female students and staff, as well as high attrition rates.

\section{Conclusions}
Our study shines a spotlight on the difficulties women experience during their doctoral studies, and the additional hurdles they have to overcome in order to succeed in their academic careers. The additional difficulties women experience during their doctoral studies provide, however, only a partial explanation to the low proportion of women in physics graduate studies and academic staff. As we learned from previous research \cite{lewis2016,lewis2017,ceci2010}, a number of cultural, social, environmental, and biological factors play a role in women's relatively lower representation in physics and other science, technology, engineering, and math (STEM) fields. Given its persistence, the causes of gender disparity are likely to be complex and multiply determined.

The gender-specific hurdles that we identified are hidden to the academic system, which is the reason that we call them {\it glass hurdles}. The academic institutes believe, almost religiously, in the ideology of meritocracy, together with a liberal concept of equality in their demands from the individual. Our research implies that, at the same time, these institutes are not taking care of leveling the playing ground, are not aware of the organizational climate in which women experience discrimination and sexual harassment, and do not deal with the gaps that are generated between female students and their male colleagues, when the former become mothers.

It is in the interest of the academic institutes and of the discipline of physics to increase the percentage of women along all the stages of an academic career in physics. Based on our findings, we suggest the following steps to be taken by academic institutes to remove the presently-transparent glass hurdles to women:
\begin{itemize}
\item Addressing  the problem of sexual harassment and promoting prevention programs.
\item Promoting inclusive teaching and discrimination-free environment for women.
\item Adapting the institutional policy to the special challenges that arise when combining Ph.D. studies with family demands.
\item Raising the fellowship for students that are parents.
\item Expanding the availability of psychological care for students in general, and female students in particular.
\end{itemize}

\subsection*{Acknowledgements}
We thank Sharon Diamant-Pick for her help in conducting the survey and in analyzing the interviews.
YN is the Amos de-Shalit chair of theoretical physics. This research is supported by grants from the Israeli Ministry of Science and Technology, from the Estate of Rene Lustig and from the Estate of Jacquelin Eckhous.
\vspace{6 pt}



\begin{thebibliography}{99}
\bibitem{footnote:UCU}
University and College Union (UCU) is a trade union and professional association for academics, lecturers, trainers, researchers and academic-related staff working in further and higher education throughout the UK.
\end{thebibliography}
\end{document}